\documentclass[secnumarabic,amssymb, nobibnotes,showpacs]{revtex4-1}

\usepackage{graphicx}	
\usepackage{amsmath}
\usepackage{subcaption}

\begin{document}

\title{Accurate determination of time delay and embedding dimension for state space reconstruction from a scalar time series}%

\author{Aniruddha T V}%
\email[]{aniruddha.venkata@cbs.ac.in}
\affiliation{UM-DAE Centre for Excellence in Basic Science, Mumbai, India}
\author{Bhaskar Lachman Khubchandani}%
\email[]{khubchandani@cbs.ac.in}
\affiliation{UM-DAE Centre for Excellence in Basic Science, Mumbai, India}

\begin{abstract}
 A new and accurate method to determine the time delay and embedding dimension for state space 
 reconstruction of a high dimensional system from a scalar time series using time delay embedding is 
 presented. The time delay is obtained to unprecedented accuracy by evaluating the minima of a newly defined 
 dimension deviation function. The efficacy of our method is tested by applying it to the Lorenz system 
 and the Mackey-Glass system. A good agreement is obtained between the shape and embedding dimension 
 of the physical system attractor(s) and the corresponding reconstruction(s) for both the systems studied. 
 This, along with a heuristic argument provide a validation of the proposed method.
 \end{abstract}
\maketitle
\section{Introduction}
Often, as a result of experimental limitations, only one dimensional data is available for 
chaotic physical systems which have higher dimensionality. The dripping faucet\cite{shaw,shaw1} 
is one such chaotic system. Another such system is the Rayleigh-Benard convective
system\cite{benard,rayleigh}, which was experimentally realised by 
Castaing et al \cite{libchaber}.
\\  The technique of state space reconstruction is used widely in analysis 
of time series data. It finds applications,for example; in analysis of the time series 
obtained from multi-filamentation in optical beams, fiber solitons and ocean rogue waves\cite{simon}.
It was concluded that predictability of rogue wave phenomenon in oceans is 
feasible in a interval of $5\tau$, where tau is the delay time determined using linear auto correlation. It is therefore, 
important to determine the delay time accurately.
It also finds applications in  analysis of chaotic data from rainfall and other 
climatic systems\cite{jayawardane}. 
Time delay techniques are often used in analysis of financial time series and 
stock trends\cite{saad}.
Therefore, accurately understanding phase space dynamics is of paramount 
importance in characterizing, predicting and eventually controlling chaos.  

 Yet another system of particular interest is the system 
in \cite{raj}, which yields a fifteen dimensional attractor having an intrinsic time delay of $22ns$ governed by 
delay differential equations used to predict the system time series upto several delay periods. 
Handling chaotic experimental data has always posed a challenge. Grassberger and 
Proccacia\cite{grassberger} defined the correlation dimension, as a scalable 
alternative to capacity and information dimension, for finite data sets. 
\\Chaotic systems such as the Lorenz system \cite{lorenz} and the Mackey-Glass system \cite{mg} yield solutions
 that lie on well characterised and multi-dimensional strange attractors \cite{wolf}. 
 The shape of these attractors and their dimensionality has also been well characterised. 
 Among the differences between the Lorenz system and the Mackey-Glass system is that the latter has a 
 well defined time-delay parameter in its governing nonlinear delay differential equation whereas the former is 
 governed by a set of three nonlinear coupled ordinary differential equations without any explicit time delay. 
 \section{The Method}
\begin{figure}[htb]
\centering
  \begin{subfigure}[b]{.24\linewidth}
    \centering
    \includegraphics[width=.99\textwidth]{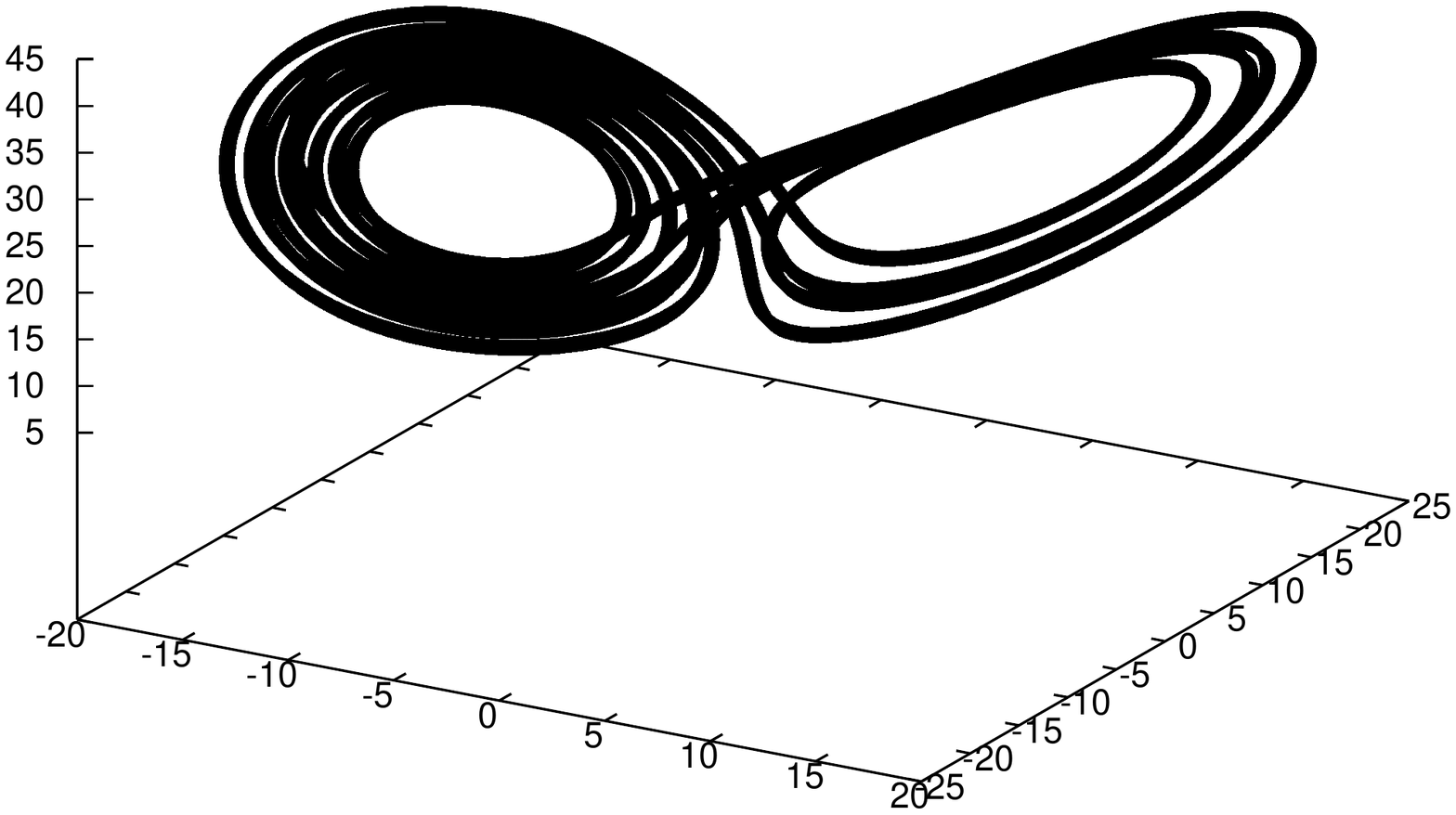}
    \caption{}
 \label{fig:1a}
  \end{subfigure}%
  \begin{subfigure}[b]{.24\linewidth}
    \centering
    \includegraphics[width=.99\textwidth]{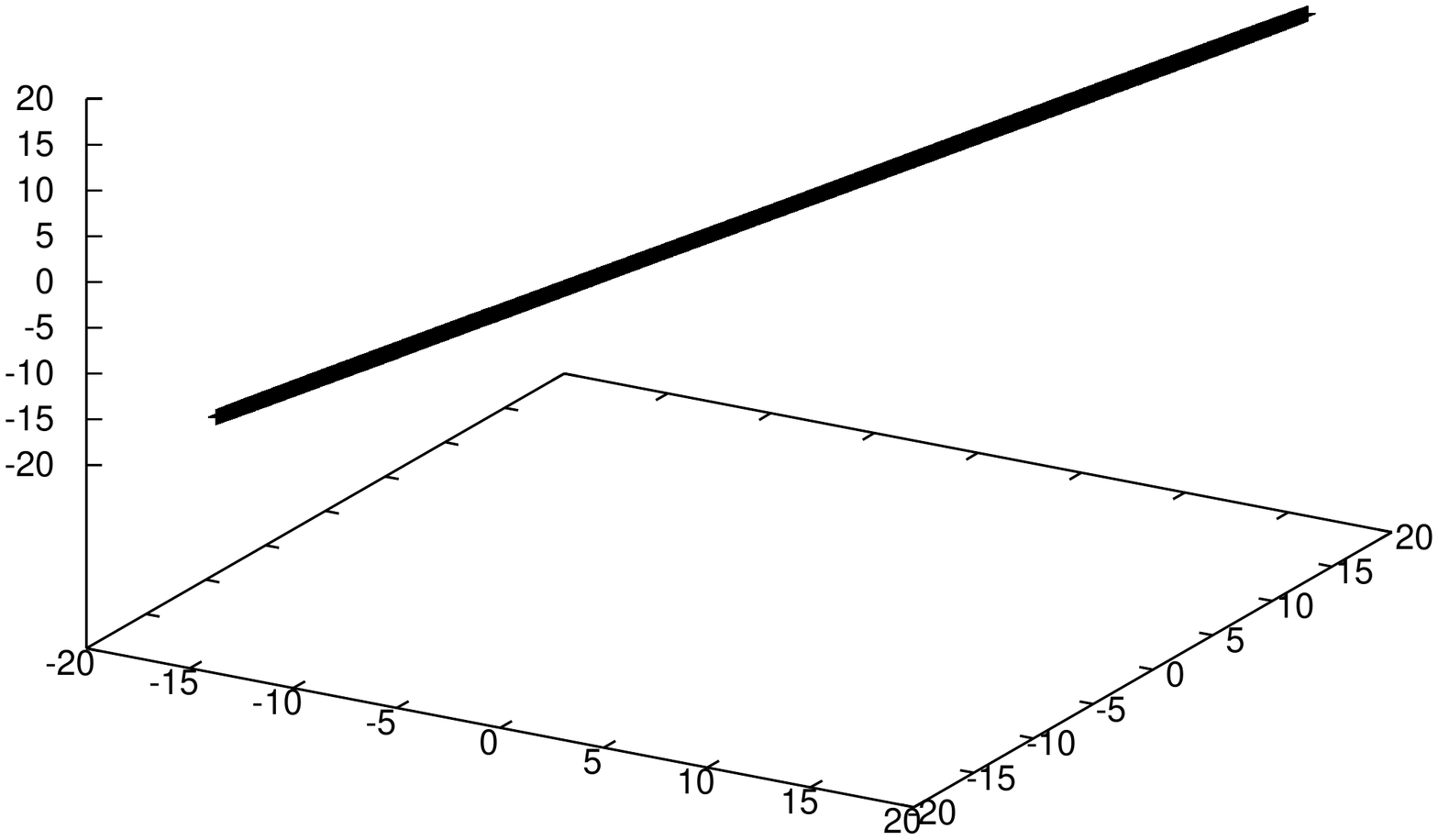}
    \caption{$\tau$=5}
    \label{fig:1b}
  \end{subfigure}%
  \begin{subfigure}[b]{.24\linewidth}
    \centering
    \includegraphics[width=.99\textwidth]{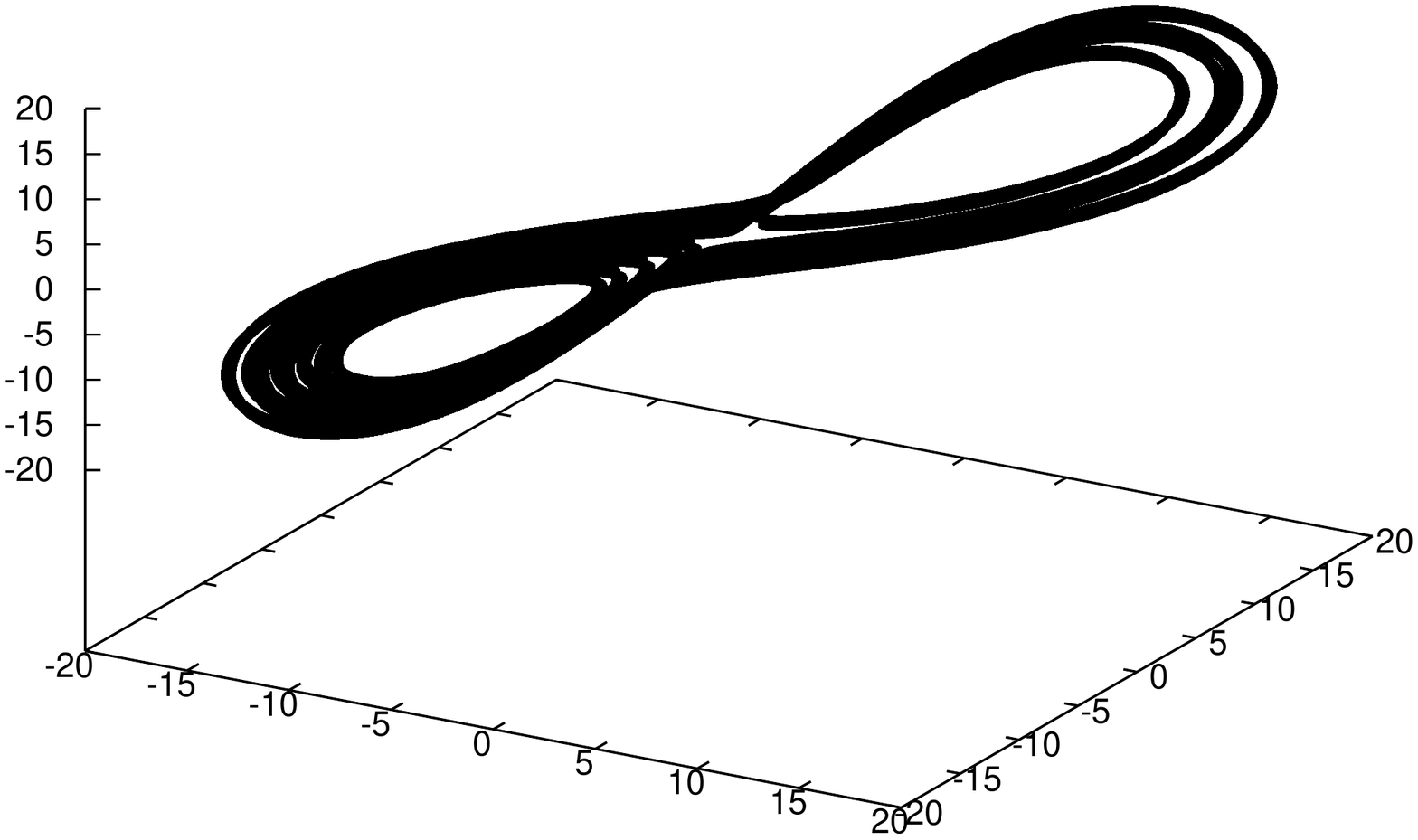}
    \caption{$\tau$=500}
    \label{fig:1c}
  \end{subfigure}%
  \begin{subfigure}[b]{.24\linewidth}
    \centering
    \includegraphics[width=.99\textwidth]{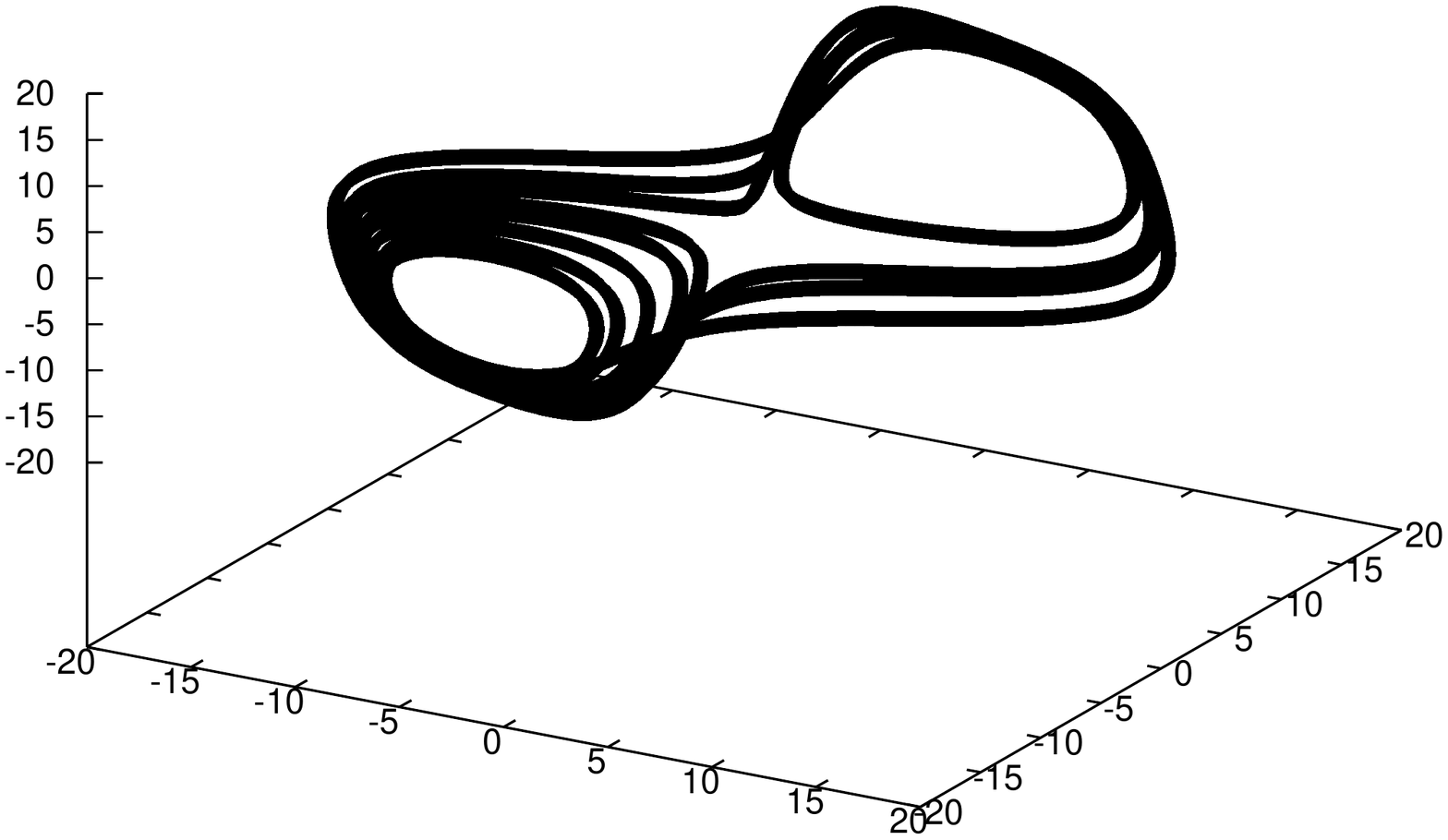}
    \caption{$\tau$=1500}
    \label{fig:1d}
  \end{subfigure}\\%
  \begin{subfigure}[b]{.24\linewidth}
    \centering
    \includegraphics[width=.99\textwidth]{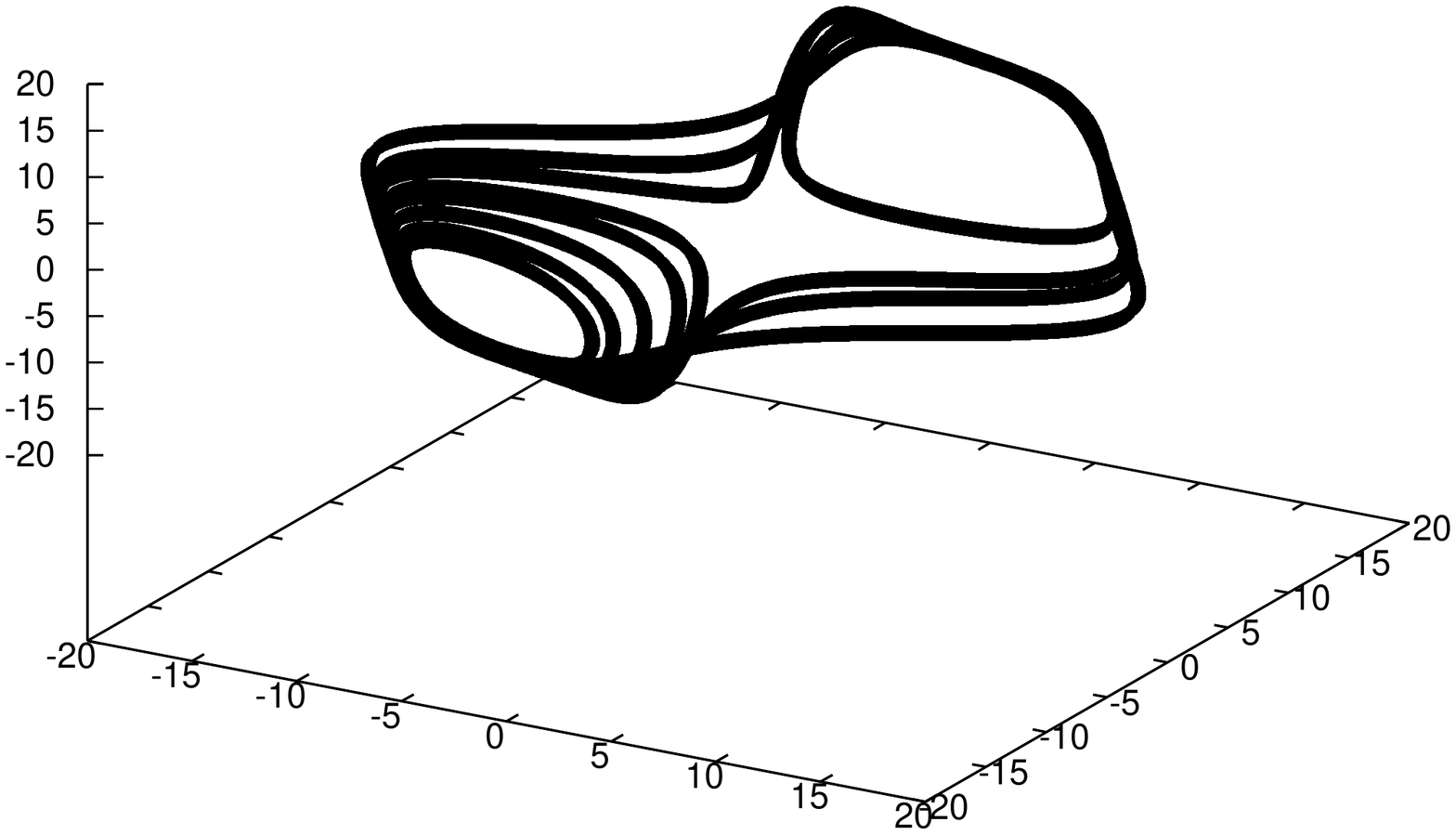}
    \caption{$\tau$=1800}
    \label{fig:1e}
  \end{subfigure}%
  \begin{subfigure}[b]{.24\linewidth}
    \centering
    \includegraphics[width=.99\textwidth]{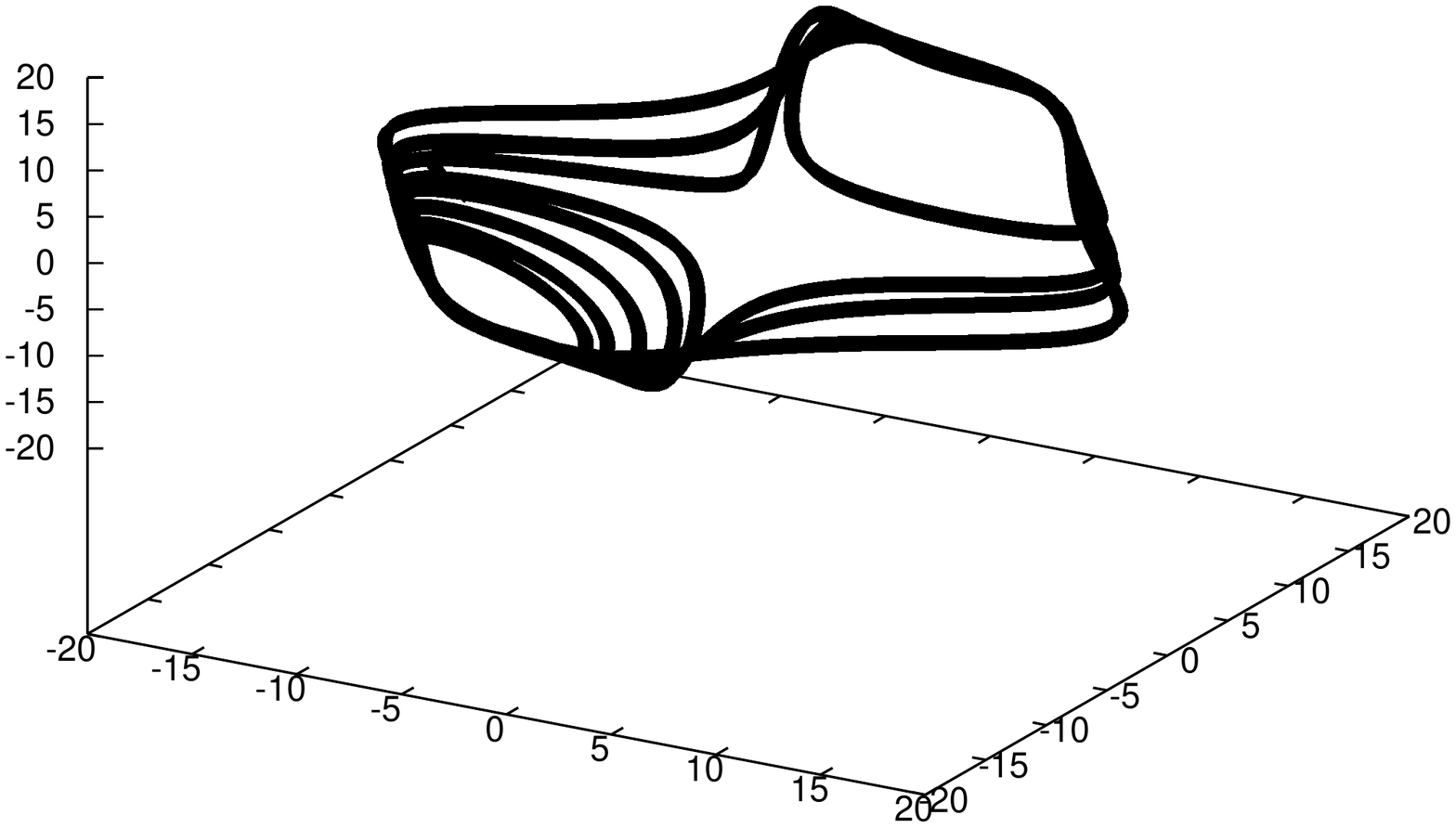}
    \caption{$\tau$=2000}
    \label{fig:1f}
  \end{subfigure}%
  \begin{subfigure}[b]{.24\linewidth}
    \centering
    \includegraphics[width=.99\textwidth]{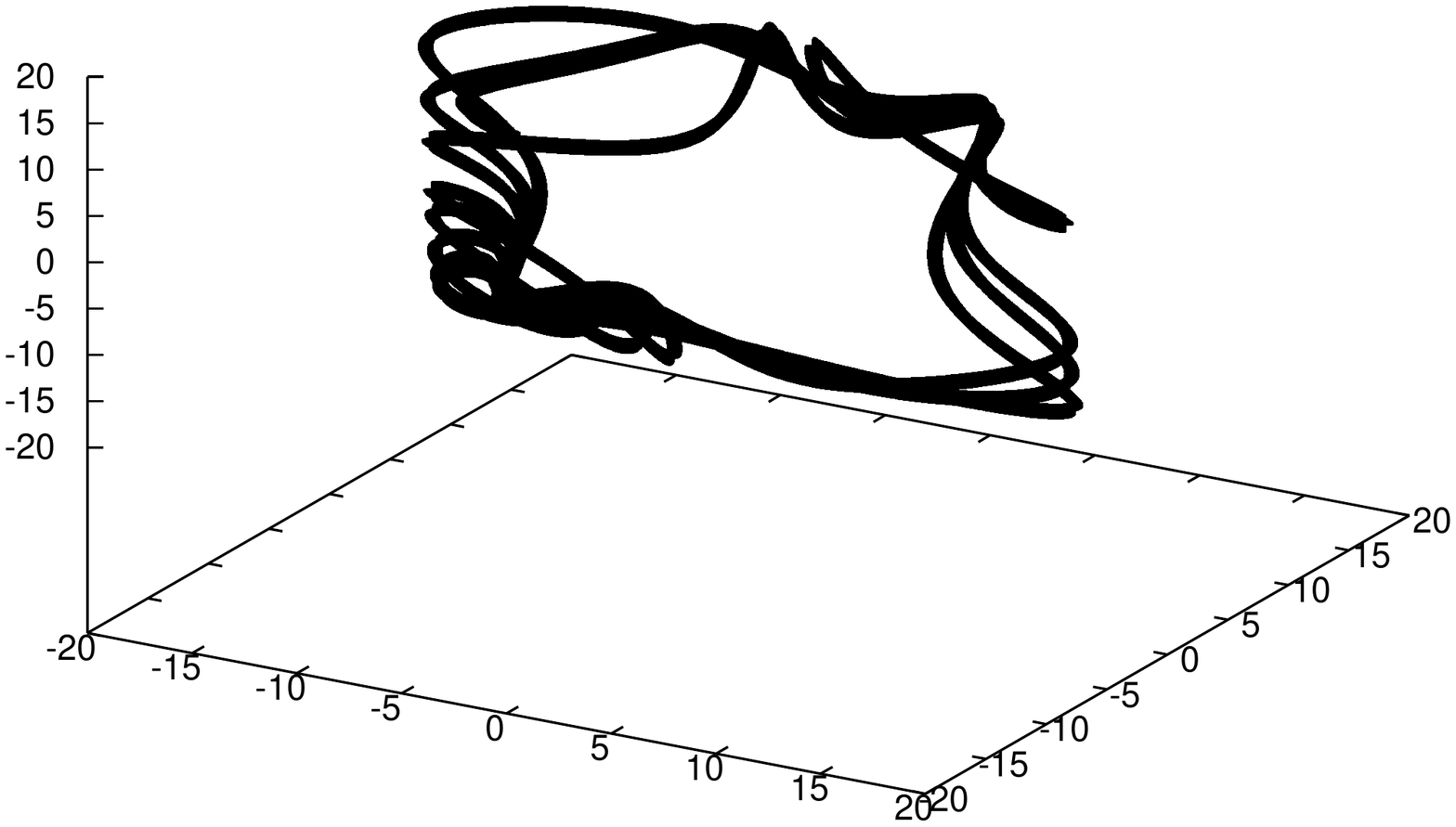}
    \caption{$\tau$=3000}
    \label{fig:1g}
  \end{subfigure}%
  \begin{subfigure}[b]{.24\linewidth}
    \centering
    \includegraphics[width=.99\textwidth]{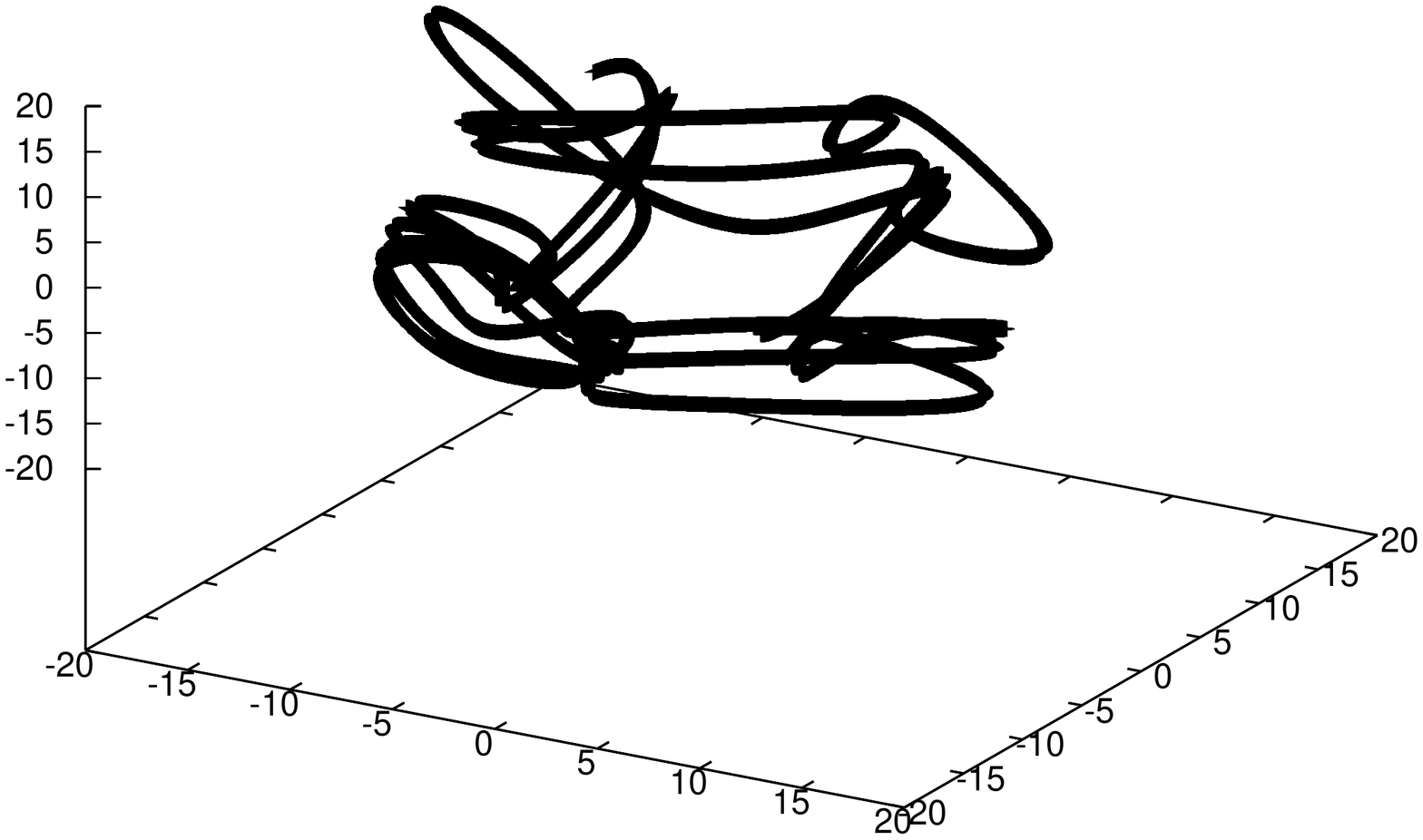}
    \caption{$\tau$=5042}
  \label{fig:1h}
  \end{subfigure}%
  
  \caption{\ref{fig:1a} shows the original Lorenz attractor 
  \ref{fig:1b}-\ref{fig:1h} show the reconstructions of the Lorenz attractor from a time series for successively 
  larger values of delay coordinate. The time series was generated by using the x coordinates of $10^5$ 
  successive points.
  The numerical solution of the Lorenz equation was obtained using the euler method from the parameters $\sigma=10$,
  $\beta=2.667$ and $\rho=28$, with initial condition $x=-10.4,y=-20.6, z=30.5$ . 
  The high degree of similarity between fig:1a and fig:1e vindicates the 
  method used for reconstruction}
  \label{fig:1}
\end{figure}

\begin{figure}[!]
 \centering
  \includegraphics[width=0.5\textwidth]{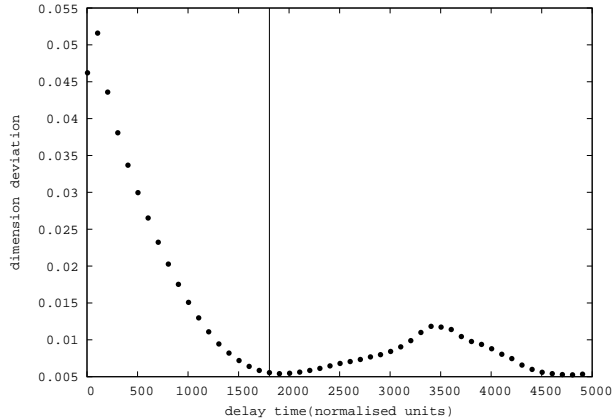}
 \caption{figure shows the dimension-deviation function versus time delay for 1-dimensional data
  from the lorenz system. The 1st
 minima was found at $\tau$=1805, approximately the same as the best reconstruction.Time is in normalised units }
 \label{fig:2a} 
 \end{figure}

In this Letter, we present a method to reconstruct the multi-dimensional state space and strange attractor 
of a chaotic system using a one-dimensional time series arrived at from solution of the governing differential equations 
without a priori knowledge of any implicit time delay. We address the problem of accurately determining time-delay 
and embedding dimension for state space reconstruction of high dimensional chaotic systems using 
one-dimensional system data. \\
The first step in this direction is the Whitney embedding theorem \cite{whitney} which states that a map 
from an n-manifold to a $2n+1$ dimensional Euclidean space is an embedding. Subsequently, Takens\cite{takens}
showed that  an n-manifold can be recovered from a single measured quantity. 
It was shown \cite{takens} that time delayed versions of the measured 
quantity [s(t),s($t+\tau$)...s($t+2n\tau$)] would embed the n-manifold. However, 
data from physical systems do not indicate a natural choice for the delay 
coordinate $\tau$ and embedding dimension $2n+1$.  
Figure \ref{fig:1} illustrates that choice of $\tau$ affects the reconstruction 
significantly. We are hence motivated to give a prescription to choose the delay 
time $\tau$ efficiently and accurately. Linear auto correlation function has popularly been used to delay time.
Further,  Fraser and Swinney \cite{swinney}
have suggested the use of average mutual information to choose the delay coordinate. However,
in our present case, we 
found that neither choice  yielded an appealing reconstruction. We hence, are 
motivated to suggest a prescription of our own. We briefly discuss the choice
of embedding dimension.  \\
A key idea that we use in this Letter is that of fractal dimension. We hence 
make some elementary definitions of importance. 
We first denote a open ball of radius $\epsilon$ centred at the point $x$, by $B_\epsilon 
(x)$. We then let $\mu(S)$ denote the natural measure associated with set S.
The point wise fractal dimension $D_p$ may be defined as \cite{Ott};
\begin{equation}
   D_p=\lim_{\epsilon \to 0} \frac{log(\mu(B_{\epsilon}(x)))}{log(\epsilon)}
\end{equation}
The remarkable feature of the point wise dimension is that it is independent of 
the point $x$ that is chosen. A heuristic argument for this may be found in Ott\cite{Ott}.
A more comprehensive review of fractal dimensions may be found in Farmer et al's 
work\cite{farmer}.
\\ We test our methodology on the well chracterized Lorenz \cite{lorenz} and Mackey-Glass \cite{mg} systems. 
The Lorenz system is given by
\begin{eqnarray}
  \frac{dx}{dt}=\sigma (y-x)
  \\ \frac{dy}{dt}=x(\rho - z)-y
  \\ \frac{dz}{dt}=xy-\beta z
  \end{eqnarray}
  The Mackey-Glass system is given by;
  \begin{equation}
 \frac{dx}{dt}=\frac{\beta x_\tau}{1+x_\tau^n}-\gamma x
 \end{equation}
 Where $x_\tau$ represents the value of $x$ at time $t-\tau$ . The values of 
 $\beta$, $\gamma$, $n$ and $\tau$ were chosen to be 2,1,10 and 1500 respectively.
\\ We now propose a new prescription for the choice of delay coordinate in a 
reconstruction. However, we would first require to make a guess for the embedding 
 dimension $m$. But, in principle, once $m$ is determined by the prescriptions 
 suggested later, the process may be repeated. 
 We let s(i) denote the $i^{th}$ entry of the time series. 
 \\
 Then we write the euclidean distances as:
 \begin{equation}
   r_{ij}=\sqrt{\sum_{k=0}^{m-1} [s(i+k\tau)-s(j+k\tau)]^2}
 \end{equation}
\\ The measure $\mu$ is;
\begin{align}
   \mu = \frac{1}{N}
   \sum_{j=1}^{N}\theta(\epsilon-r_{ij} )
   \label{dp1}
 \end{align}
  where $\theta$ is the unit step function, N the total number of state space points 
  obtained, and $\epsilon$, an arbitrarily small number. It is then easy to see that the following 
  equation for $D_p$ would correspond to the expression for the pointwise 
  dimension taken centred about point ``$i$"
  \begin{equation}
    D_p(i,\tau)=\lim_{\epsilon \to 0} \frac{log(\mu(i,\tau,\epsilon))}{log(\epsilon)}
    \label{dp}
  \end{equation} 
  We next define the dimension deviation function, f as,
  \begin{equation}
    f(\tau)=\frac{1}{N}\sum_{i=1}^{N} (D_p(i,\tau)-\bar{D_p})^2
    \label{8}
  \end{equation}
  where $\bar{D_p}$ is the value of the pointwise dimension averaged over all 
  points.
  \begin{equation}
    \bar{D_p}=\frac{1}{N}\sum_{i=1}^{N} D_p(i,\tau)
  \end{equation}
  We now claim that the minima of $f(\tau)$ is a good choice for $\tau$. \\ We 
  motivate this claim, with the following argument. We first notice that$\mu(i,\tau,\epsilon)$
  is the measure associated with an open ball or radius $\epsilon$ centred around 
  a point in phase space, labelled $i$, reconstructed with delay time $\tau$. From 
  Equation \ref{dp}, it follows that $D_p(i,\tau)$ is the pointwise dimension 
  of the attractor. Hence $f(\tau)$ is the standard deviation in the pointwise 
  dimension. Should the reconstruction be a one that recovers most of the 
  attracting set dynamics, we would expect to obtain zero standard deviation (since the pointwise dimension
  is invariant with respect to the point chosen, for an attracting set \cite{Ott}). Hence a minima 
  in the standard deviation would definitely occur if the attractor is fully recovered, since the 
  standard deviation is necessarily a positive quantity. \\ \\ We may further argue that only if there exists a set
  of points ${S_1}$ where the measure remains invariant and positive and another disjoint set $S_2$ with another
  value of the measure would we see non zero or relatively larger values of standard deviation 
  in the pointwise dimension. However the measure on the set $S_1 \cup S_2$ would not be 
  ergodic, and hence cannot correspond to an attracting set of a smooth map.
  An ergodic measure $\mu$ cannot be decomposed into two measures, $\mu1$ 
  and $\mu2$, such that \cite{Ott};
  \begin{equation}
    \mu=p\mu_1+(1-p)\mu_2
  \end{equation}
  Where p is any real which lies in the interval (0,1).
  \\ It may however prove tricky to actually compute the pointwise dimension 
 from a finite quantity of data, since no finite amount of data can give an 
 accurate estimate of measure. It is therefore recommended to use very small 
 values of $\epsilon$ in Equation \ref{dp}, but allow only those values that contain 
 at least two points within the open ball, to avoid outliers. A regression fit of 
 $log(\epsilon)$ against $log(\mu)$ ought to give a reliable estimate of the the pointwise dimension.
 Further, computations can be cut down by choosing to use a large representative 
 set of points, as the centers for the computation of the pointwise dimension, rather than the entire data set.
 \\ In the case of the Lorenz attractor, we observe that the fist local minima of the dimension deviation
 function obtained at $\tau=1805$ gives the best reconstruction observed visually. Fig \ref{fig:2a} shows the dimension deviation
 as a function of $\tau$
  
  \begin{figure}[!]
    \centering
    \includegraphics[width=0.5\textwidth]{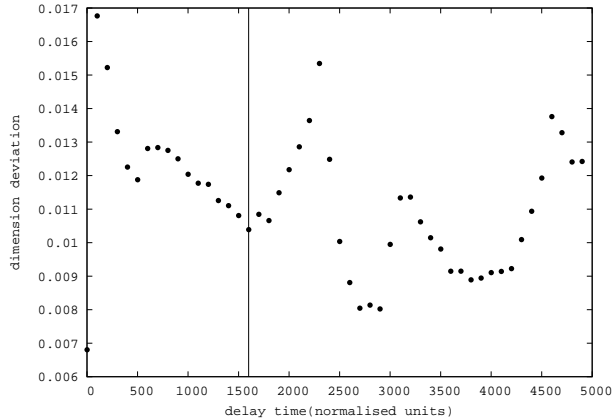}
    \caption{Figure shows
 The dimesion-deviation function against delay time for the Mackey-Glass attractor that was studied. The first 
 minima was found at $\tau$=1600. The delay coordinate chosen in the underlying 
 delay differential equation is $\tau=1500$. Time is in normalised units.}
    \label{fig:2b}
 \end{figure} 
   Figure \ref{fig:2b} shows dimension deviation function for the Mackey-Glass attractor.  
   The first minima was obtained at $\tau=1605$ units, while the delay used in 
   the underlying attractor was $\tau=1500$ units. This high degree of accuracy 
   indicates that  the first local minima of the dimension deviation is indeed a
   good choice for the delay coordinate.
   \section{Other methods}
 Some other methods for determining the optimal delay have been proposed in 
literature\cite{swinney}\cite{Abarbanel}. The first method uses the linear auto correlation function.
The linear auto correlation is defined  by the following 
   relationship.
   \begin{equation}
     C_l(\tau)=\frac{\frac{1}{N}\sum_{m=1}^{N}[s(m+\tau)-\bar{s}][s(m)-\bar{s}]}{\frac{1}{N}[s(m)-\bar{s}]^2}
   \end{equation}
   where $\bar{s}$ is the average value of the time series. We immediately 
   remark that this definition follows from finding the best fit function $C_l(\tau)$ 
   for the linear relationship,
   \begin{equation}
     s(n+\tau)-\bar{s}=C_l(\tau)[s(n)-\bar{s}]
   \end{equation}
   The prescription often used for the choice of $\tau$ is the first zero of the
   auto-correlation function defined above \cite{Abarbanel}.
    It is easy to see that, the linear auto-correlation function, may 
    yield a bad choice for $\tau$ for nonlinear systems, since minimising the linear dependance of terms separated
    by a time-span of $\tau$, does not necessarily minimise the over all dependance that arises from 
    the non linear terms. Further, minimising the dependance of terms separated by
    $\tau$ may not be the best strategy, since we are looking for an intermediate $\tau$
    such that terms separated by a distance of $\tau$ are neither statistically 
    independent, nor nearly overlapping.
    \\ In the study of the Lorenz attractor, we found that that the auto 
    correlation had its first zero, far from the point where the best visual 
    reconstruction was found. This demonstrates the failure of this prescription
    for nonlinear systems. 
    Yet another prescription, used often is the mutual information function. It is 
 a generalization of the linear auto-correlation function, and relates the 
 information content in one set to the information content in another. It was 
 first proposed by Gallaghar\cite{Gallagar}. In the context of  time series analysis, we 
 measure the mutual information content, of terms separated by a distance 
 $\tau$.We define mutual information between terms separated by a distance $\tau$ 
 to be;
 \begin{equation}
   I(\tau)=\sum_{n}P[n,n+\tau]log(\frac{P[n] P[n+\tau]}{P[n,n+\tau]})
 \end{equation}
 where $P[n]$ is the probability measure for the occurrence of s(n) 
  and $P[n,n+\tau]$ represents the joint probability of their occurrence.
The prescription often suggested is the use of the first minima of the  average mutual information
 as an appropriate choice of $\tau$ \cite{swinney}.
 However, since $I(\tau)$ also measures the information between two terms 
 separated, by $\tau$, we expect that its first minima is close to the zero of the linear 
 auto-correlation function.
 $P(s(n))$ in determined by the relative frequency of the occurrence of the value $s(n)$ and 
 $P(s(n+\tau))$ is determined likewise. $P(s(n),s(n+\tau))$ is determined by the 
 relative frequency of occurrence of the the pair of numbers $s(n)$ and 
 $s(n+\tau)$, separated by exactly a time-span of $\tau$.
 \\ In the present study of the Lorenz attractor,  the plot of the average mutual information against $\tau$
  yielded a minima that was far from the delay time used in the best visual reconstruction. 
  However, it was closer to the  optimal value of delay time, as compared to that predicted by the linear 
  auto-correlation.
 Hence, this method too yields a value that is far off in the present case.
\section{Embedding dimension} 
 While there exist many methods \cite{kennan,broomhead,glass} that one may use to determine the optimal value 
 of the embedding dimension, we suggest one that is along the lines of the method of 
 false nearest neighbours listed in Abarbanel et al's \cite{Abarbanel} work. We rewrite Equation \ref{dp1}, 
 however now making it a function of $m$, keeping $\tau$ fixed.
 \begin{equation}
      \nu (i,m,\epsilon)= 
   \sum_{j=1}^{N}\theta(\epsilon-r_{ij})
 \end{equation}
Here $\nu$ indicates the total number of nearest neighbours.
 We now observe that at a low dimension the number of false neighbours at every 
 point would be higher. However, when embedded in any dimensionality higher than 
 the optimal embedding dimension, the number of neighbours would remain nearly 
 the same. Hence looking for the point of saturation of the total number of neighbours, against the 
 embedding dimension, would give us a good estimate of the optimal embedding dimension. 
 
  In the present case, for the Lorenz attractor, using  the plot of embedding dimension against the number of neighbours
  it was found to saturate at a embedding dimension value of 5. 
\section{Summary}
 To summarise, we have tested the methods for delay coordinate choice given in literature and found that they 
 have not succeeded in our case study of the Lorenz and Mackey-Glass systems. We further developed an 
 alternative prescription for the choice 
 of delay coordinate, modelled after the deviation in the pointwise dimension. It worked significantly better for 
 our particular case. 
 \\  We then used the same definition to write down a 
 prescription for the choice in embedding dimension as well. The major shortcoming of the proposed methodology lies
in an arbitrary initial choice in embedding dimension that has to be made to 
accurately
determine the delay time. To circumvent this,  an arbitrary and high 
choice of the embedding dimension can be made to determine the optimal delay, then determine the optimal 
embedding dimension and further redo the calculation for $\tau$ in the new embedding dimension.
\\  Our method has a time complexity of 
 $O(n^2)$, while mutual information, has a 
 algorithm with time complexity $O(nlogn)$\cite{swinney}. 
However, one may average over a representative set of points rather than the 
whole set and obtain the standard deviation to reduce computation time by any 
desirable factor.

\end{document}